\documentstyle[12pt,psfig]{article}
\def\baselinestretch{1.5}
\begin{document}
\pagenumbering{arabic}
 \vskip -1.cm
 \begin{flushright}
 \fbox{BIHEP-TH ------ 2002-49}
 \end{flushright}
\begin{center}
{\Large \bf Effect of  tensor couplings in a relativistic \\
Hartree approach  for finite nuclei}\\
 \vspace{0.5cm}
Guangjun~Mao \\
\vspace{1.0cm}
{\em $^{1)}$Institute of High Energy Physics, Chinese Academy of Science \\
      P.O. Box 918(4), Beijing 100039, P.R. China\footnote{mailing address\\
      \indent $\;\;$ e-mail: maogj@mail.ihep.ac.cn}\\
$^{2)}$Institute of Theoretical Physics, Chinese Academy of Science\\
      P.O. Box 2735, Beijing 100080, P.R. China \\
$^{3)}$CCAST (World Lab.), P.O. Box 8730, Beijing 100080, P.R.
       China }
\end{center}
\date{\today}
\vspace{0.5cm}
\begin{abstract}
\begin{sloppypar}
The relativistic Hartree approach describing the bound states of
both nucleons and anti-nucleons in finite nuclei has been extended
to include tensor couplings for the $\omega$- and $\rho$-meson.
After readjusting the parameters of the model to the properties of
spherical nuclei, the effect of tensor-coupling terms rises the
spin-orbit force by a factor of 2, while a large effective nucleon
mass $m^{*}/M_{N} \approx 0.8$ sustains. The overall nucleon
spectra of shell-model states are  improved evidently. The
predicted anti-nucleon spectra in the vacuum are deepened about 20
-- 30 MeV.

\end{sloppypar}
\bigskip
\noindent {\bf PACS} number(s): 21.10.-k; 21.60.-n; 13.75.Cs
\end{abstract}
\newcounter{cms}
\setlength{\unitlength}{1mm}
\newpage
\begin{center}
{\bf I. INTRODUCTION}
\end{center}
\begin{sloppypar}
One of the main characters distinguishing  relativistic approaches
from  nonrelativistic approaches is that the former one has a
vacuum. It is quite interesting to study the structure of quantum
vacuum in a many-body system, e.g., in a finite nucleus where the
Fermi sea is filled with the valence nucleons while the Dirac sea
is full of the nucleon--anti-nucleon pairs. In the relativistic
treatment of nuclear phenomena \cite{Ser86}, the Dirac equation is
used to describe the behavior of nucleons in nuclei. The effects
of the nuclear medium on nucleons are taken into account through
introducing strong Lorentz scalar ($S$) and time-component Lorentz
vector ($V$) potential. In the language of meson-exchange theory
the scalar potential can be attributed to the exchange of sigma
meson and the vector potential to the exchange of omega and rho
mesons as well as the electromagnetic force. Since the Dirac
equation describes the nucleon and the anti-nucleon
simultaneously, the effects of mean fields act on both of them.
Consequently, not only the valence nucleons are bounded in the
shell-model like states, but also there exist bound states for
anti-nucleons emerging from the lower continuum. The observation
of anti-nucleon bound states is a verification for the application
of the Dirac phenomenology to a relativistic many-body system
\cite{Aue86}. It constitutes a basis for the widely used
relativistic mean-field (RMF) theory
\cite{Ser86,Rei86,Gam90,Ren96,Sug94,Rut97} and the relativistic
Hartree approach (RHA) \cite{Hor84,Per86,Was88,Fox89,Fur89}. Since
the bound states of nucleons are subject to the cancellation of
two potentials $S+V$ ($V$ is positive, $S$ is negative) while the
bound states of anti-nucleons, due to the G-parity, are sensitive
to the sum of them $S-V$, consistent studies of both the nucleon
and the anti-nucleon bound states can determine the individual $S$
and $V$. In addition, the exact knowledge of potential depth for
anti-nucleons in the medium is a prerequisite for the study of
anti-matter and anti-nuclei in relativistic heavy-ion collisions
\cite{Bea00,Arm00}.

The shell-model states have been theoretically and experimentally
well established \cite{Boh69} while no information for the bound
states of anti-nucleons in the Dirac sea are available. This is
the aim of our work. In Ref. \cite{Mao99} we have developed a
relativistic Hartree approach which describes the bound states of
nucleons and anti-nucleons consistently. The contributions of the
Dirac sea to the source terms of meson-field equations are
considered up to the one-nucleon loop and one-meson loop and
evaluated by means of the derivative expansion technique
\cite{Ait84}. The parameters of the model are adjusted by fitting
to the properties of spherical nuclei. The major outcome of the
RHA model is that a rather large effective nucleon mass
$m^{*}/M_{N} \approx 0.8$ is obtained compared to the value of 0.6
in the relativistic mean-field calculations where the {\it no-sea}
approximation is adopted . This is caused by the effects of the
vacuum contributions which decrease the magnitude of the scalar
potential $S$ substantially. Correspondingly, the vector potential
$V$ is also suppressed since the quantity $V+S$ is controlled by
the saturation properties. As pointed out above, the anti-nucleon
bound states are mainly determined by the sum of the scalar and
vector potentials $S-V$. The smaller values of $S$ and $V$
obtained in the RHA calculations lead to a weaker bound on the
single-particle energies of anti-nucleons, which turn out to be
only half of that computed in the RMF model. On the other hand,
the spin-orbit potential of nucleons is related to $d(S-V)/dr$. In
the RHA calculations the spin-orbit splitting of the shell-model
states is roughly 1/3 of that calculated in the RMF approach and
indicated by the empirical data, although the general trend of the
energy spectra coincides with each other. Because our goal is to
develop a model to predict the bound states of anti-nucleons in
the vacuum  with the model parameters constrained by the nuclear
bulk properties, in order to get reliable results for the
anti-nucleon spectra one should first describe the nucleon spectra
as good as possible.

Theoretically one can incorporate tensor couplings for the
$\omega$- and $\rho$-meson, which mainly contribute to the
spin-orbit force. The model now becomes non-renormalizable.
However, from the point of view of modern effective field theory
\cite{Wei95} the argument of renormalizability is not a severe
restriction to theories. Since we will employ an effective
Lagrangian of mesonic degrees of freedom and nucleons, just as the
Skyrme force is an effective Lagrangian for nonrelativistic
calculations \cite{Fri86}, the mesons here are effective mesons.
In an effective field theory one normally evaluates the
lowest-order diagrams and makes regularization whenever a
divergence appears. The parameters of the Lagrangian are adjusted
to fit certain experimental data. The validity of the whole
approach is justified by successful explanations and predictions
of observables. In other words, within the framework of an
effective field theory one mainly concerns the balance between the
predictive ability of the theory and the complexity of the theory.

In the present work we will investigate the effects of tensor
couplings in the relativistic Hartree approach for finite nuclei.
In this extended version of the model the parameters will be
rearranged in a least-square fit to the properties of spherical
nuclei. The model is then applied to study the bound states of
nucleons and anti-nucleons. The paper is organized as follows: In
Sect. II we introduce the effective Lagrangian and review the RHA
model. In Sect. III we present the numerical results and
discussions. A summary and outlook are finally given in Sect. IV.

\end{sloppypar}
\begin{center}
{\bf II. RELATIVISTIC HARTREE APPROACH}
\end{center}
\begin{sloppypar}
The Lagrangian density of nucleons interacting through the
exchange of mesons can be expressed as \cite{Ser86}
\begin{equation}
{\cal L}={\cal L}_{\rm F}+{\cal L}_{\rm I}.
\end{equation}
Here ${\cal L}_{F}$ is the Lagrangian density for free nucleon,
mesons and photon
\begin{eqnarray}
{\cal L}_{\rm
F}&=&\bar{\psi}[i\gamma_{\mu}\partial^{\mu}-M_{N}]\psi
   + \frac{1}{2}
\partial_{\mu}\sigma\partial^{\mu}\sigma-U(\sigma)
 -\frac{1}{4}\omega_{\mu\nu}\omega^{\mu\nu} \nonumber \\
&& + \frac{1}{2}m_{\omega}^{2}\omega_{\mu}\omega^{\mu}
 - \frac{1}{4} {\bf R}_{\mu\nu} \cdot {\bf R}^{\mu\nu}
 +\frac{1}{2}m_{\rho}^{2}{\bf R}_{\mu} \cdot {\bf R}^{\mu}
- \frac{1}{4}  A_{\mu\nu} A^{\mu\nu}
    \end{eqnarray}
and U($\sigma$) is the self-interaction part of the scalar field
\cite{Bog77}
\begin{equation}
  U(\sigma)=
   \frac{1}{2}m_{\sigma}^{2}\sigma^{2}+\frac{1}{3!}b
\sigma^{3}+\frac{1}{4!}c\sigma^{4}.
\end{equation}
In the above expressions $\psi$ is the Dirac spinor of the
nucleon; $\sigma$, $\omega_{\mu}$, ${\bf R}_{\mu}$ and $A_{\mu}$
represent the scalar meson, vector meson, isovector-vector meson
field and the electromagnetic field, respectively. Here the field
tensors for the omega, rho and photon are given in terms of their
potentials by
 \begin{eqnarray}
&&  \omega_{\mu\nu}=\partial_{\mu}\omega_{\nu}
   -\partial_{\nu}\omega_{\mu},\\
&&  {\bf R}_{\mu\nu}=\partial_{\mu}{\bf R}_{\nu}
-\partial_{\nu}{\bf
R}_{\mu}, \\
&&   A_{\mu\nu}=\partial_{\mu}A_{\nu} -\partial_{\nu}A_{\mu}.
  \end{eqnarray}
${\cal L}_{I}$ is the interaction Lagrangian density
\begin{eqnarray}
 {\cal L}_{I}&=&{\rm g}_{\sigma}\bar{\psi}\psi\sigma
      - {\rm g}_{\omega}\bar{\psi}\gamma_{\mu}\psi\omega^{\mu}
      -\frac{f_{\omega}}{4M_{N}}\bar{\psi}\sigma^{\mu\nu}\psi\omega_{\mu\nu}
  - \frac{1}{2}{\rm g}_{\rho}\bar{\psi}\gamma_{\mu}\mbox{\boldmath $\tau$}
    \cdot \psi {\bf R}^{\mu} \nonumber \\
 && -\frac{f_{\rho}}{8M_{N}}\bar{\psi}\sigma^{\mu\nu}\mbox{\boldmath
$\tau$}\cdot \psi {\bf R}_{\mu\nu} - \frac{1}{2}
e\bar{\psi}(1+\tau_{0})\gamma_{\mu} \psi A^{\mu}.
\end{eqnarray}
Here $\sigma_{\mu\nu}=\frac{i}{2}\left[
\gamma_{\mu},\gamma_{\nu}\right]$, $\mbox{\boldmath $\tau$}$ is
the isospin operator of the
 nucleon and $\tau_{0}$ is its third component. ${\rm g}_{\sigma}$, ${\rm
g}_{\omega}$, ${\rm g}_{\rho}$ and $e^{2}/4\pi = 1/137$ are the
coupling strengths for the $\sigma$-, $\omega$-, $\rho$-meson  and
for the photon, respectively. $f_{\omega}$ and $f_{\rho}$ are the
tensor-coupling strengths of vector mesons.  $M_{N}$ is the free
nucleon mass and $m_{\sigma}$, $m_{\omega}$, $m_{\rho}$ are the
masses of the $\sigma$-, $\omega$-, and $\rho$-meson.

In finite nuclei the Dirac equation is written as
 \begin{eqnarray}
 i \frac{\partial}{\partial t}\psi({\bf x},t) &=&
        \left[ -i \mbox{\boldmath $\alpha$} \cdot \mbox{\boldmath $\nabla$}
        + \beta\left(M_{N} - {\rm g}_{\sigma}\sigma({\bf x})\right)
        +{\rm g}_{\omega}\omega_{0}({\bf x})
        -\frac{f_{\omega}}{2M_{N}}i\mbox{\boldmath $\gamma$}\cdot
        \left(\mbox{\boldmath $\nabla$} \omega_{0}({\bf x})\right)
        \right. \nonumber \\
    &+& \left. \frac{1}{2}{\rm g}_{\rho}\tau_{0}R_{0,0}({\bf x})
        -\frac{f_{\rho}}{4M_{N}}i\tau_{0}\mbox{\boldmath $\gamma$}\cdot
        \left(\mbox{\boldmath $\nabla$}R_{0,0}({\bf x})\right)
        +\frac{1}{2}e(1+\tau_{0})A_{0}({\bf x})\right] \psi({\bf x},t).
        \label{field}
 \end{eqnarray}
The field operator can be expanded according to nucleons and
anti-nucleons and reads as
 \begin{equation}
\psi({\bf x},t)=\sum_{\alpha} \left[ b_{\alpha}\psi_{\alpha}({\bf
 x}) e^{-i E_{\alpha}t} + d^{+}_{\alpha}\psi^{a}_{\alpha}({\bf x})
 e^{i \bar{E} _{\alpha} t} \right]. \label{opera}
  \end{equation}
Here the label $\alpha$ denotes the full set of single-particle
quantum numbers. $\psi_{\alpha}({\bf x})$ are the wave functions
of nucleons and $\psi_{\alpha}^{a}({\bf x})$ are those of
anti-nucleons; $E_{\alpha}$ and $\bar{E}_{\alpha}$ are their
positive energies, respectively. $b_{\alpha}$ and $d^{+}_{\alpha}$
are the annihilation and creation operators of nucleons and
anti-nucleons that satisfy the standard anticommutation relations.
We assume that the meson fields depend only on the radius and
discuss the problem in spherically symmetric nuclei. In this case,
the usual angular momentum and parity are good quantum numbers. As
described in Refs. \cite{Bjo64,Gre89}, the eigenfunctions of
nucleons are the well-known spherical spinors
  \begin{equation}
 \psi_{\alpha}({\bf x})= \left( \begin{array}{l}
 i \frac{G_{\alpha}(r)}{r} \Omega_{jlm}(\frac{{\bf r}}{r})  \\
  \frac{F_{\alpha}(r)}{r}\frac{\mbox{\boldmath $\sigma$}\cdot {\bf r}}{r}
 \Omega_{jlm}(\frac{{\bf r}}{r}) \end{array} \right). \label{wfn}
  \end{equation}
We make the ansatz for the wave functions of anti-nucleons
\cite{Mao99}
  \begin{equation}
 \psi_{\alpha}^{a}({\bf x})= \left( \begin{array}{l}
 \frac{\bar{F}_{\alpha}(r)}{r}\frac{\mbox{\boldmath $\sigma$}\cdot {\bf r}}{r}
 \Omega_{jlm}(\frac{{\bf r}}{r})  \\
  i \frac{\bar{G}_{\alpha}(r)}{r} \Omega_{jlm}(\frac{{\bf r}}{r})
 \end{array} \right). \label{wfan}
  \end{equation}
Here $\Omega_{jlm}$ are the spherical spinors defined as
  \begin{equation}
\Omega_{jlm}=\sum_{m^{\prime}m_{s}}\left( l \frac{1}{2} j \mid
m^{\prime}
 m_{s} m \right) Y_{l m^{\prime}} \chi_{\frac{1}{2}m_{s}},
  \end{equation}
$Y_{l m^{\prime}}$ are the spherical harmonics and
$\chi_{\frac{1}{2}m_{s}}$ are the eigenfunctions of the spin
operators. $G_{\alpha}$, $F_{\alpha}$ and
 $\bar{F}_{\alpha}$, $\bar{G}_{\alpha}$ are the remaining {\em real} radial
wave functions of nucleons and anti-nucleons for upper and lower
components, respectively.

Inserting Eq. (\ref{opera}) into Eq. (\ref{field}) one immediately
obtains two relativistic wave equations for the
$\psi_{\alpha}({\bf x})$ and $\psi_{\alpha}^{a}({\bf x})$.
Applying the concrete expressions of the wave functions given in
(\ref{wfn}) and (\ref{wfan}), we arrive at the coupled equations
for the radial wave functions of nucleons
  \begin{eqnarray}
E_{\alpha}G_{\alpha}(r)&=&\left[-\frac{d}{dr}+\frac{\kappa_{\alpha}}{r}
   -\frac{f_{\omega}}{2M_{N}}\left(\partial_{r}\omega_{0}(r)\right)
   -\frac{f_{\rho}}{4M_{N}}\tau_{0\alpha}\left(\partial_{r}R_{0,0}(r)\right)
  \right]  F_{\alpha}(r) \nonumber \\
   && + \left[ M_{N} - {\rm g}_{\sigma}\sigma(r) + {\rm g}_{\omega}
     \omega_{0}(r) +\frac{1}{2}{\rm g}_{\rho}\tau_{0\alpha}R_{0,0}(r)
   ¡¡\right. ¡¡\nonumber \\
&&  \left.  +\frac{1}{2}e\left(1+\tau_{0\alpha}\right)A_{0}(r)
   \right] G_{\alpha}(r), \\
E_{\alpha}F_{\alpha}(r)&=&\left[\frac{d}{dr}+\frac{\kappa_{\alpha}}{r}
  -\frac{f_{\omega}}{2M_{N}}\left(\partial_{r}\omega_{0}(r)\right)
  -\frac{f_{\rho}}{4M_{N}}\tau_{0\alpha}\left(\partial_{r}R_{0,0}(r)\right)
  \right]G_{\alpha}(r) \nonumber \\
  && +  \left[ - M_{N} + {\rm g}_{\sigma}\sigma(r) + {\rm g}_{\omega}
   \omega_{0}(r) +\frac{1}{2}{\rm g}_{\rho}\tau_{0\alpha}R_{0,0}(r)
   \right. \nonumber \\
&& \left.  +\frac{1}{2}e\left(1+\tau_{0\alpha}\right)A_{0}(r)
   \right] F_{\alpha}(r)
  \end{eqnarray}
and anti-nucleons
  \begin{eqnarray}
 -\bar{E}_{\alpha}\bar{F}_{\alpha}(r)&=&\left[\frac{d}{dr}
 +\frac{\kappa_{\alpha}}{r}+\frac{f_{\omega}}{2M_{N}}\left(\partial_{r}
 \omega_{0}(r)\right) + \frac{f_{\rho}}{4M_{N}}\bar{\tau}_{0\alpha}
 \left(\partial_{r}R_{0,0}(r)\right)\right]\bar{G}_{\alpha}(r)\nonumber \\
& & + \left[ M_{N} - {\rm g}_{\sigma}\sigma(r) + {\rm g}_{\omega}
   \omega_{0}(r)+\frac{1}{2}{\rm g}_{\rho}\bar{\tau}_{0\alpha}R_{0,0}(r)
      \right. \nonumber \\
 && \left.  +\frac{1}{2}e\left(1-\bar{\tau}_{0\alpha}\right)A_{0}(r)
   \right] \bar{F}_{\alpha}(r), \\
 -\bar{E}_{\alpha}\bar{G}_{\alpha}(r)&=&\left[ - \frac{d}{dr} +
  \frac{\kappa_{\alpha}}{r} +\frac{f_{\omega}}{2M_{N}}\left(\partial_{r}\omega_{0}
  (r)\right) +\frac{f_{\rho}}{4M_{N}}\bar{\tau}_{0\alpha}\left(\partial_{r}
  R_{0,0}(r)\right) \right]\bar{F}_{\alpha}(r) \nonumber \\
& & + \left[ -M_{N} + {\rm g}_{\sigma}\sigma(r) + {\rm g}_{\omega}
 \omega_{0}(r)+\frac{1}{2}{\rm g}_{\rho}\bar{\tau}_{0\alpha}R_{0,0}(r)
   \right. \nonumber \\
 && \left. +\frac{1}{2}e\left(1-\bar{\tau}_{0\alpha}\right)A_{0}(r) \right]
 \bar{G}_{\alpha}(r),
  \end{eqnarray}
  where
  \begin{equation}
\kappa_{\alpha}= \left\{  \begin{array}{ll}
-(l+1) & \mbox{for $j=l+\frac{1}{2}$} \\
 l & \mbox{for $j=l-\frac{1}{2}$} \end{array} \right.
  \end{equation}
and $\bar{\tau}_{0\alpha}$ is the isospin factor of anti-nucleons,
$\bar{\tau}_{0\alpha}=-\tau_{0\alpha}$. In the numerical solution
of the relativistic wave equations one eliminates the small
components to obtain the Schr\"{o}dinger-equivalent equations. For
the nucleon we eliminate the lower component while for the
anti-nucleon the upper component. By defining the
Schr\"{o}dinger-equivalent effective mass and potentials of  the
nucleon
 \begin{eqnarray}
&&  M_{eff}= E_{\alpha}+ M_{N} - {\rm g}_{\sigma}\sigma (r)
 - {\rm g}_{\omega}\omega_{0}(r) -\frac{1}{2}{\rm
 g}_{\rho}\tau_{0\alpha}R_{0,0}(r) -\frac{1}{2}e
 \left(1+\tau_{0\alpha}\right) A_{0}(r), \label{effmn} \\
&& U_{eff}= M_{N} - {\rm g}_{\sigma}\sigma(r)
 + {\rm g}_{\omega}\omega_{0}(r)+\frac{1}{2}{\rm
 g}_{\rho}\tau_{0\alpha}R_{0,0}(r) + \frac{1}{2}e\left(
 1+\tau_{0\alpha}\right) A_{0}(r), \\
 && W(r)=\frac{\kappa_{\alpha}}{r}-\frac{f_{\omega}}{2M_{N}}\left(
    \partial_{r}\omega_{0}(r)\right)
    -\frac{f_{\rho}}{4M_{N}}\tau_{0\alpha}\left(\partial_{r}R_{0,0}(r)\right)
 \end{eqnarray}
and the anti-nucleon
 \begin{eqnarray}
&&  \bar{M}_{eff}= \bar{E}_{\alpha}+
    M_{N}-{\rm g}_{\sigma}\sigma(r)
 + {\rm g}_{\omega}\omega_{0}(r) -\frac{1}{2}{\rm
 g}_{\rho}\tau_{0\alpha}R_{0,0}(r) +\frac{1}{2}e
 \left(1+\tau_{0\alpha}\right) A_{0}(r), \\
&& \bar{U}_{eff}= M_{N} - {\rm g}_{\sigma}\sigma(r)
 - {\rm g}_{\omega}\omega_{0}(r)+\frac{1}{2}{\rm
 g}_{\rho}\tau_{0\alpha}R_{0,0}(r) - \frac{1}{2}e\left(
 1+\tau_{0\alpha}\right) A_{0}(r), \\
 && \bar{W}(r)=\frac{\kappa_{\alpha}}{r}+\frac{f_{\omega}}{2M_{N}}\left(
    \partial_{r}\omega_{0}(r)\right)
    -\frac{f_{\rho}}{4M_{N}}\tau_{0\alpha}\left(\partial_{r}R_{0,0}(r)\right),
  \label{wan}
 \end{eqnarray}
we arrive at the Schr\"{o}dinger-equivalent  equations for the
upper component of the nucleon's  wave function
 \begin{equation}
E_{\alpha}G_{\alpha}(r) = \left[- \frac{d}{dr} + W(r)
 \right] M_{eff}^{-1} \left[ \frac{d}{dr} + W(r) \right]
 G_{\alpha}(r) + U_{eff}G_{\alpha}(r) \label{schn}
 \end{equation}
and the lower component of the anti-nucleon's wave function
 \begin{equation}
\bar{E}_{\alpha}\bar{G}_{\alpha}(r) = \left[-
\frac{d}{dr}+\bar{W}(r) \right] \bar{M}_{eff}^{-1} \left[
\frac{d}{dr} + \bar{W}(r) \right]
 \bar{G}_{\alpha}(r) + \bar{U}_{eff}\bar{G}_{\alpha}(r).
 \label{schan}
 \end{equation}
The small components can be obtained through the following
relations
 \begin{eqnarray}
&& F_{\alpha}(r) = M_{eff}^{-1} \left[ \frac{d}{dr} + W(r)
   \right] G_{\alpha}(r), \\
&& \bar{F}_{\alpha}(r) = - \bar{M}_{eff}^{-1} \left[ \frac{d}{dr}
   + \bar{W}(r)\right] \bar{G}_{\alpha}(r).
 \end{eqnarray}
 In the above we have changed $\bar{\tau}_{0\alpha} \rightarrow
 -\tau_{0\alpha}$, i.e., now the anti-nucleon has the same isospin
 factor as the corresponding nucleon.
From Eqs. (\ref{schn}) and (\ref{schan}) one finds that the
Schr\"{o}dinger equation of the anti-nucleon has the same form as
that of the nucleon. The only difference relies on the definition
of the effective mass and potentials, that is, the vector fields
change their signs. The so-called {\em G-parity} comes out
automatically.

It's now a suitable place to discuss the orthonormalization
condition of the wave functions. In Eq. (\ref{opera}) the
annihilation and creation operators of nucleons and anti-nucleons
satisfy the usual anticommutation relations. From the equal-time
anticommutation relation of the Dirac field operator one can
derive the matrix equation for the normalization of the wave
functions
\begin{equation}
\sum_{\alpha}\left[ \psi_{\alpha}({\bf x},t)\psi_{\alpha}^{+}({\bf
y},t)+\psi_{\alpha}^{a}({\bf x},t)\psi_{\alpha}^{a+}({\bf y},t)
\right]=\delta({\bf x}-{\bf y}). \label{norm}
\end{equation}
In static case one can eliminate the $t$ index. The following
orthogonal conditions can be obtained from the Dirac wave
equations \cite{Wei95}
\begin{eqnarray}
&& \int\, d^{3}x\psi_{\alpha}^{+}({\bf x})\psi_{\beta}({\bf x})=0
   \;\;\;\;\;\;\;\;\;\;\;\;\; {\rm if} \;\;\; \alpha \neq \beta, \\
&& \int\,d^{3}x\psi_{\alpha}^{a+}
    ({\bf x})\psi_{\beta}^{a}({\bf x})=0
   \;\;\;\;\;\;\;\;\;\;\; {\rm if} \;\;\; \alpha \neq \beta, \\
&&\int\,d^{3}x\psi_{\alpha}^{+}
   ({\bf x})\psi_{\beta}^{a}({\bf x})=0.\label{org}
\end{eqnarray}
Multiplying Eq. (\ref{norm}) on the right with $\psi_{\alpha}({\bf
y})$ or $\psi_{\alpha}^{a}({\bf y})$, it can be found that these
wave functions must satisfy the normalization conditions
\begin{equation}
\int\, d^{3}y \psi_{\beta}^{+}({\bf y})\psi_{\alpha}({\bf y})
=\int\, d^{3}y \psi_{\beta}^{a+}({\bf y})\psi_{\alpha}^{a} ({\bf
y})=\delta_{\alpha\beta}.
\end{equation}
This leads to the orthonormalization conditions for the radial
wave functions of nucleons and anti-nucleons
\begin{eqnarray}
\int_{0}^{\infty}\, dr \left[ G_{\alpha}(r)G_{\beta}(r) +
F_{\alpha}(r) F_{\beta}(r) \right] =\delta_{\alpha\beta} , \\
\int_{0}^{\infty}\, dr \left[\bar{G}_{\alpha}(r)\bar{G}_{\beta}(r)
+\bar{F}_{\alpha}(r) \bar{F}_{\beta}(r) \right]
=\delta_{\alpha\beta},
\end{eqnarray}
respectively. The single-particle energies of the nucleon and the
anti-nucleon can be evaluated as
 \begin{eqnarray}
 E_{\alpha} &=& \int^{\infty}_{0} dr \lbrace G_{\alpha}(r) \left[
 -\frac{d}{dr} + W(r) \right] F_{\alpha}(r) + F_{\alpha}(r)
 \left[ \frac{d}{dr} + W(r) \right] G_{\alpha}(r)
 \nonumber \\
&&  + G_{\alpha}(r)U_{eff}G_{\alpha}(r) - F_{\alpha}(r) \left[
 M_{eff}- E_{\alpha} \right] F_{\alpha}(r) \rbrace, \\
 \vspace{0.2cm}
 \bar{E}_{\alpha} &=& \int^{\infty}_{0} dr \lbrace - \bar{F}_{\alpha}(r)
 \left[\frac{d}{dr} + \bar{W}(r) \right]\bar{G}_{\alpha}(r)
 - \bar{G}_{\alpha}(r)
 \left[ -\frac{d}{dr} + \bar{W}(r) \right] \bar{F}_{\alpha}(r)
  \nonumber \\
&&  + \bar{G}_{\alpha}(r)\bar{U}_{eff}\bar{G}_{\alpha}(r)
 - \bar{F}_{\alpha}(r) \left[ \bar{M}_{eff}
 - \bar{E}_{\alpha} \right] \bar{F}_{\alpha}(r) \rbrace,
 \end{eqnarray}
which are obtained through the iteration procedure.

The main ingredients in Eqs. (\ref{effmn}) - (\ref{wan}) are the
meson fields, which are determined by the Laplace equations
 \begin{eqnarray}
&& \left( \mbox{\boldmath $\nabla$}^{2} - m_{\sigma}^{2} \right)
 \sigma(r) = - {\rm g}_{\sigma} \rho_{S}(r) +
 \frac{1}{2}b\sigma^{2}(r) + \frac{1}{3!}c \sigma^{3}(r), \label{eqsig} \\
&& \left( \mbox{\boldmath $\nabla$}^{2} - m_{\omega}^{2} \right)
 \omega_{0} (r) = - {\rm g}_{\omega} \rho_{0} (r)
  -\frac{f_{\omega}}{2M_{N}}\rho_{0}^{T}(r), \\
  && \left( \mbox{\boldmath $\nabla$}^{2} - m_{\rho}^{2}\right)R_{0,0}(r)=
 - \frac{1}{2}{\rm g}_{\rho}\rho_{0,0}(r)
  -\frac{f_{\rho}}{4M_{N}}\rho_{0,0}^{T}(r) , \\
&& \mbox{\boldmath $\nabla$}^{2} A_{0}(r) = -e
   \rho_{Pr,0}(r),\label{eqele}
 \end{eqnarray}
 and
 \begin{equation}
\mbox{\boldmath
$\nabla$}^{2}=\frac{d^{2}}{dr^{2}}+\frac{2}{r}\frac{d}{dr}
 \end{equation}
in the case of spherical nuclei. The source terms of the
meson-field equations are various densities which, in principle,
should contain the contributions both from the Fermi sea and the
Dirac sea. Under the mean-field approximation, those densities are
formally computed as the expectation values of various bilinear
products of field operators in the ground state of the many-body
system. For example, a direct calculation of the baryon density
gives
\begin{eqnarray}
\rho_{0} & = & \langle \psi_{0} \mid \psi^{+}\psi \mid \psi_{0}
         \rangle \nonumber \\
         &=& \sum_{\alpha=1}^{A}\psi_{\alpha}^{+}\psi_{\alpha}
         +\sum_{\alpha=vac}\psi_{\alpha}^{a+}\psi_{\alpha}^{a},
\end{eqnarray}
where the sum on the second term of the last equality runs over
all anti-nucleon spectra in the vacuum, and therefore causes
divergence. A proper regularization scheme is apparently needed in
order to render it to a finite value. Unfortunately, at the
situation of finite nuclei it is currently untractable. If one
simply cuts off it, the integration of the term with respect to
the space doesn't vanish. This violets the baryon number
conservation. In an alternative point of view, the mean fields are
taken as a starting point for calculating corrections within the
framework of Quantum Hadrodynamics \cite{Ser97}. That is, one
drops the second term and includes quantum corrections by means of
Feynman diagrams and path-integral methods. Within this scheme,
the contributions of the valence nucleons to the densities are
computed by adding up wave functions while the contributions of
the Dirac sea are taken into account in loop expansions. At the
one-loop level the effective action of the system can be written
as \cite{Jac74}
  \begin{eqnarray}
\Gamma &=& \int\, d^{4}x \left(\frac{1}{2}
\partial_{\mu}\sigma \partial^{\mu}\sigma
   -U(\sigma) - \frac{1}{4}\omega_{\mu\nu}\omega^{\mu\nu} + \frac{1}{2}m_{\omega}
   ^{2}\omega_{\mu}\omega^{\mu} -\frac{1}{4}{\bf R}_{\mu\nu}\cdot
   {\bf R}^{\mu\nu} + \frac{1}{2}m_{\rho}^{2}{\bf R}_{\mu}\cdot{\bf
    R}^{\mu} \right.  \nonumber \\
   && \left. -\frac{1}{4}A_{\mu\nu}A^{\mu\nu} + CT \right) + \Gamma_{{\rm
   valence}}+ \frac{i}{2}\hbar {\rm Tr}\ln \left( iD^{-1} \right)
 -i\hbar {\rm Tr}\ln \left(iG^{-1}\right) . \label{effact}
  \end{eqnarray}
Here the $CT$ are the counterterms. $\Gamma_{{\rm valence}}$ is
the contribution from the valence nucleons, which for
time-independent background fields is just minus the energy of the
valence nucleons. The last two terms in Eq. (\ref{effact})
represent the contributions of the Dirac sea stemming from the
one-meson loop and one-nucleon loop, respectively. By means of the
derivative expansion technique \cite{Ait84} they can be expressed
as
\begin{eqnarray}
\Gamma^{(1)}(\sigma)&=& \frac{i}{2}\hbar {\rm Tr}\ln
             \left(iD^{-1}\right) \nonumber \\
      &=& \int\, d^{4}x \left(-V_{B}^{(1)}(\sigma)+\frac{1}{2}
 Z^{(1)}(\sigma)(\partial_{\mu} \sigma)^{2} + ... \right), \label{g1s} \\
 \Gamma^{(1)}(\psi)&=&-i\hbar {\rm Tr}\ln \left(iG^{-1}\right)
                  \nonumber \\
      &=& \int\, d^{4}x \left( - V_{F}^{(1)}(\sigma) +\frac{1}{2}
 Z_{1\sigma}^{(1)}(\sigma)(\partial_{\mu}\sigma)^{2} + \frac{1}{4}Z^{(1)}
 _{1\omega}(\sigma) \omega_{\mu\nu}\omega^{\mu\nu} \right. \nonumber \\
   && \left. +\frac{1}{4}Z_{1A}^{(1)}(\sigma)A_{\mu\nu}A^{\mu\nu}+ ... \right).
 \end{eqnarray}
It can be verified that the tensor-coupling terms contribute to
the higher-order terms in the derivative expansion and have
therefore been neglected. $V_{B}^{(1)}(\sigma)$ and
$V_{F}^{(1)}(\sigma)$ are the effective potentials from the
one-meson loop and one-nucleon loop, in which the field is a
constant, $\sigma(x)=\sigma_{0}$, the same situation as in nuclear
matter. These two terms contain divergent part and should be
regularized. Through adding suitable counterterms,
$V_{B}^{(1)}(\sigma)$ and $V_{F}^{(1)}
 (\sigma)$ can be calculated in nuclear matter which turn out to be
\cite{Lee74,Chi77,Ser86}
  \begin{eqnarray}
V_{B}^{(1)}(\sigma)&=& \frac{m_{\sigma}^{4}}{(8\pi)^{2}} \left[
       \left( 1 + \frac{b\sigma}{m_{\sigma}^{2}} +
       \frac{c\sigma^{2}}{2m_{\sigma}^{2}} \right)^{2}
       \ln \left( 1+ \frac{b\sigma}{m_{\sigma}^{2}} +
      \frac{c\sigma^{2}} {2m_{\sigma}^{2}} \right) - \left(
      \frac{b\sigma}{m_{\sigma}^{2}} + \frac{
      c\sigma^{2}}{2m_{\sigma}^{2}} \right) \right. \nonumber \\
 && \left.
 -\frac{3}{2} \left( \frac{b\sigma}{m_{\sigma}^{2}} + \frac{c\sigma^{2}}
 {2m_{\sigma}^{2}} \right)^{2} - \frac{1}{3} \left( \frac{b\sigma}{m_{\sigma}
 ^{2}} \right) ^{2} \left( \frac{b\sigma}{m_{\sigma}^{2}} + \frac{3c\sigma^{2}}
 {2m_{\sigma}^{2}} \right) + \frac{1}{12} \left( \frac{b\sigma}{m_{\sigma}^{2}}
 \right) ^{4} \right], \\
V_{F}^{(1)}(\sigma)&=& - \frac{1}{4\pi^{2}} \left[ \left( M_{N} -
     {\rm g}_{\sigma}\sigma \right)^{4}\ln \left( 1- \frac{{\rm
     g}_{\sigma}\sigma} {M_{N}} \right) + M_{N}^{3}{\rm
     g}_{\sigma}\sigma - \frac{7}{2}M_{N}^{2}
      {\rm g}_{\sigma}^{2}\sigma^{2} \right. \nonumber \\
 && \left. + \frac{13}{3}M_{N}{\rm g}_{\sigma}^{3}
     \sigma^{3} - \frac{25}{12}{\rm g}_{\sigma}^{4}\sigma^{4} \right].
  \end{eqnarray}
The functional coefficients before various derivative terms can be
determined in the derivative expansion technique and read as
\cite{Per86,Fur89}
 \begin{eqnarray}
&&Z^{(1)}(\sigma)=\frac{1}{12 }\frac{(b+c\sigma)^{2}}
 {16\pi^{2}(m_{\sigma}^{2} + b\sigma
 +\frac{1}{2}c\sigma^{2}) }, \\
&& Z^{(1)}_{1\sigma}(\sigma)= - \frac{{\rm
g}_{\sigma}^{2}}{2\pi^{2}} \ln
 \left( \frac{m^{*}}{M_{N}} \right), \\
&& Z^{(1)}_{1\omega}(\sigma)=  \frac{{\rm
g}_{\omega}^{2}}{3\pi^{2}} \ln
 \left( \frac{m^{*}}{M_{N}} \right),\\
 && Z^{(1)}_{1A}(\sigma)=  \frac{e^{2}}{6\pi^{2}} \ln
 \left( \frac{m^{*}}{M_{N}} \right). \label{z1a}
  \end{eqnarray}
Inserting Eqs. (\ref{g1s}) - (\ref{z1a}) into Eq. (\ref{effact})
and minimizing the effective action with respect to the
corresponding fields, one reveals the meson-field equations as
given before, in which $\Gamma_{\rm valence}$ constitutes the
densities originated from the valence nucleons while
$\Gamma^{(1)}(\sigma)$ and $\Gamma^{(1)}(\psi)$ compose the
densities stemming from the Dirac sea. At the end we obtain
concrete expressions of various densities contained in Eqs.
(\ref{eqsig}) -- (\ref{eqele})
 \begin{eqnarray}
 && \rho_{S}(r)=\rho_{S}^{val}(r)+\rho_{S}^{sea}(r), \\
&& \rho_{0}(r)=\rho_{0}^{val}(r)+\rho_{0}^{sea}(r), \\
&& \rho_{0}^{T}(r) = \rho_{0}^{T,val}(r), \\
&& \rho_{0,0}(r) = \rho_{0,0}^{val}(r), \\
&& \rho_{0,0}^{T}(r) = \rho_{0,0}^{T,val}(r), \\
&& \rho_{Pr,0}(r)=\rho_{Pr,0}^{val}(r) + \rho_{Pr,0}^{sea}(r),
 \end{eqnarray}
and
 \begin{eqnarray}
 \rho_{S}^{val}(r)&=&\frac{1}{4\pi r^{2}}\sum_{\alpha=1}^{\Omega}
      w_{\alpha}(2 j_{\alpha}+ 1)\left[G_{\alpha}^{2}(r) -
      F_{\alpha}^{2}(r)\right],  \\
\rho_{0}^{val}(r)&=&\frac{1}{4\pi r^{2}}\sum_{\alpha=1}^{\Omega}
      w_{\alpha}(2 j_{\alpha}+ 1)\left[G_{\alpha}^{2}(r) +
      F_{\alpha}^{2}(r)\right],  \\
\rho_{0}^{T,val}(r)&=&\frac{1}{4\pi r^{2}}\sum_{\alpha=1}^{\Omega}
      w_{\alpha}(2 j_{\alpha}+ 1)2\left[\partial_{r}G_{\alpha}(r)
      F_{\alpha}(r)\right],  \\
\rho_{0,0}^{val}(r)&=&\frac{1}{4\pi r^{2}}\sum_{\alpha=1}^{\Omega}
      w_{\alpha}(2 j_{\alpha}+ 1)\tau_{0\alpha}\left[G_{\alpha}^{2}(r) +
      F_{\alpha}^{2}(r)\right],  \\
\rho_{0,0}^{T,val}(r)&=&\frac{1}{4\pi
      r^{2}}\sum_{\alpha=1}^{\Omega}
      w_{\alpha}(2 j_{\alpha}+ 1)\tau_{0\alpha}2\left[\partial_{r}G_{\alpha}(r)
      F_{\alpha}(r)\right],  \\
\rho_{Pr,0}^{val}(r) & =& \frac{1}{2}\left[ \rho_{0}^{val}(r)
      + \rho_{0,0}^{val}(r) \right], \\
\rho_{S}^{sea}(r)&=& -\frac{1}{{\rm
      g}_{\sigma}}\frac{\partial}{\partial \sigma}\left[
      V_{B}^{(1)}(\sigma)+V_{F}^{(1)}(\sigma)\right] +\frac{1}{2{\rm g}_{\sigma}}
      \left[\frac{\partial}{\partial\sigma} Z^{(1)}(\sigma) \right]
      \left[ \frac{d}{dr}\sigma(r)\right] ^{2} \nonumber \\
  &&  + \frac{1}{{\rm g}_{\sigma}}Z^{(1)}(\sigma) \left(\frac{d^{2}}{dr^{2}}
      + \frac{2}{r}\frac{d}{dr} \right) \sigma(r) + \frac{{\rm
      g}_{\sigma}^{2}}{4\pi^{2} m^{*}} \left[\frac{d}{dr}\sigma(r) \right]^{2}
      \nonumber \\
  &&  -\frac{{\rm g}_{\sigma}}{2\pi^{2}}\ln \left(\frac{m^{*}}{M_{N}}\right)
      \left( \frac{d^{2}}{dr^{2}}+\frac{2}{r}\frac{d}{dr} \right)
      \sigma(r) + \frac{{\rm g}_{\omega}^{2}}{6\pi^{2}m^{*}}
      \left[ \frac{d}{dr}\omega_{0}(r) \right] ^{2} \nonumber \\
  &&  + \frac{e^{2}}{12\pi^{2}m^{*}}\left[ \frac{d}{dr}A_{0}(r)
      \right]^{2} , \\
 \rho_{0}^{sea}(r) & =& -\frac{{\rm g}_{\omega}}{3\pi^{2}}\ln
      \left( \frac{m^{*}}{M_{N}} \right) \left(\frac{d^{2}}{dr^{2}} +
      \frac{2}{r}\frac{d}{dr} \right) \omega_{0}(r)
   + \frac{{\rm g}_{\sigma}{\rm g}_{\omega}}{3\pi^{2}m^{*}}
      \left[\frac{d}{dr}\sigma(r) \right] \left[
      \frac{d}{dr}\omega_{0}(r) \right] , \\
 \rho_{Pr,0}^{sea}(r) & =& -\frac{e}{6\pi^{2}}\ln
      \left( \frac{m^{*}}{M_{N}} \right) \left(\frac{d^{2}}{dr^{2}} +
      \frac{2}{r}\frac{d}{dr} \right) A_{0}(r)
     + \frac{e{\rm g}_{\sigma}}{6\pi^{2}m^{*}}
      \left[\frac{d}{dr}\sigma(r) \right] \left[
      \frac{d}{dr}A_{0}(r) \right] , \\
 \end{eqnarray}
where
  \begin{eqnarray}
\frac{\partial V_{B}^{(1)}(\sigma)}{\partial \sigma} &=&
\frac{m_{\sigma}^{4}}{(8\pi)^{2}} \left[ 2\left( 1 +
 \frac{b\sigma}{m_{\sigma}^{2}} + \frac{c\sigma^{2}}{2m_{\sigma}^{2}} \right)
 \left( \frac{b}{m_{\sigma}^{2}} + \frac{c\sigma}{m_{\sigma}^{2}} \right)
 \ln \left( 1 + \frac{b\sigma}{m_{\sigma}^{2}} + \frac{c\sigma^{2}}{2m_{\sigma}
 ^{2}} \right) \right. \nonumber \\
 && \left. -2 \left( \frac{b\sigma}{m_{\sigma}^{2}} + \frac{c\sigma^{2}}{2m_{\sigma}
 ^{2}} \right) \left( \frac{b}{m_{\sigma}^{2}} + \frac{c\sigma}{m_{\sigma}^{2}}
 \right) - \frac{b^{2}}{m_{\sigma}^{6}} \left( b\sigma^{2} + 2c\sigma^{3}
 \right) + \frac{b^{4}}{3 m_{\sigma}^{8}}\sigma^{3} \right], \\
\frac{\partial V_{F}^{(1)}(\sigma)}{\partial \sigma} &=&
 - \frac{1}{4\pi^{2}} \left[ - {\rm g}_{\sigma} \left( M_{N} - {\rm g}_{\sigma}
 \sigma \right) ^{3} \left( 1 + 4\ln (1-\frac{{\rm g}_{\sigma}\sigma}
 {M_{N}}) \right) + M_{N}^{3}{\rm g}_{\sigma} - 7M_{N}^{2}{\rm g}_{\sigma}^{2}
 \sigma \right. \nonumber \\
 && \left. + 13M_{N}{\rm g}_{\sigma}^{3}\sigma^{2} - \frac{25}{3}{\rm g}
 _{\sigma}^{4}\sigma^{3} \right]
\end{eqnarray}
and
 \begin{equation}
 \frac{\partial Z^{(1)}(\sigma)}{\partial \sigma}=\frac{1}{192 \pi^{2}}
 \left[ \frac{2c (b+c\sigma)}{(m_{\sigma}^{2}+ b\sigma + \frac{1}{2}c\sigma^{2})
 } - \frac{( b+c\sigma) ^{3}} {(m_{\sigma}^{2} + b\sigma + \frac{1}{2}c \sigma
 ^{2}) ^{2} } \right].
 \end{equation}
 In order to be able to calculate unclosed-shell nuclei, the
 occupation number $w_{\alpha}$ have been explicitly indicated.
Note that $\rho_{0}^{sea}(r)$ is a total derivative and thus the
baryon number is conserved. The total energy of the system can be
written as
  \begin{equation}
 E= E_{\rm MFT} + \Delta E ,
  \end{equation}
where
 \begin{eqnarray}
E_{\rm MFT} & =& \sum_{\alpha=1}^{\Omega}w_{\alpha}E_{\alpha}
     -\frac{1}{2}\int\, d^{3}x \left[ -{\rm g}_{\sigma}\sigma\rho_{S}
     + \frac{1}{6}b\sigma^{3}+ \frac{1}{12}c\sigma^{4} + {\rm g}_{\omega}
     \omega_{0}\rho_{0} \right. \nonumber \\
 &&  \left. + \frac{f_{\omega}}{2M_{N}}\omega_{0}\rho_{0}^{T}
    +\frac{1}{2}{\rm g}_{\rho}R_{0,0}\rho_{0,0} + \frac{f_{\rho}}{4M_{N}}
    R_{0,0}\rho_{0,0}^{T} + eA_{0}\rho_{Pr,0} \right], \\
 \Delta E &=& \int\, d^{3}x \left[ V_{B}^{(1)}(\sigma) + V_{F}^{(1)}(\sigma)
    + \frac{1}{2}Z^{(1)}(\sigma)(\mbox{\boldmath $\nabla$}\sigma)^{2}
    -\frac{{\rm g}_{\sigma}^{2}}{4\pi^{2}}\ln \left(\frac{m^{*}}{M_{N}}\right)
    (\mbox{\boldmath $\nabla$}\sigma)^{2} \right. \nonumber \\
 && \left. + \frac{{\rm g}_{\omega}^{2}}{6\pi^{2}}\ln \left(\frac{m^{*}}{M_{N}}
   \right) (\mbox{\boldmath $\nabla$}\omega_{0})^{2} + \frac{e^{2}}{12\pi^{2}}
   \ln \left( \frac{m^{*}}{M_{N}}\right) (\mbox{\boldmath $\nabla$}A_{0} )^{2}
   \right] .
 \end{eqnarray}
 The pairing energy and the center-of-mass correction to the total
 energy are taken into account as elucidated in Ref. \cite{Rei86}.
The energy density of homogeneous nuclear matter can be obtained
through reducing the above formulae. The compressibility at
saturation density reads as
 \begin{eqnarray}
\frac{1}{9}K &= &\frac{{\rm
     g}_{\omega}^{2}}{m_{\omega}^{2}}\rho_{0} +\frac{k_{F}^{2}}
    {3(k_{F}^{2}+m^{*2})^{1/2}} +\frac{m^{*2}\rho_{0}}{(k_{F}^{2} + m^{*2} )}
      \nonumber \\
  && \times   \left[ -\frac{1}{{\rm g}_{\sigma}^{2}}\frac{\partial^{2}}
    {\partial \sigma^{2}} \left( U(\sigma)
  +V_{B}^{(1)}(\sigma) + V_{F}^{(1)}(\sigma)
    \right)_{\sigma=\sigma_{0}}+\frac{3\rho_{0}}{(k_{F}^{2}+m^{*2})^{1/2}}
    -\frac{3\rho_{S}^{val}}{m^{*}} \right] ^{-1},
 \end{eqnarray}
 where
 \begin{equation}
\rho_{S}^{val}=\frac{4}{(2\pi)^{3}}\int_{0}^{k_{F}}\, d^{3}k
   \frac{m^{*}}{({\bf k}^{2}+m^{*2})^{1/2}}.
 \end{equation}
The double derivatives in the above expression can be easily
computed.

\end{sloppypar}
\begin{center}
{\bf III. NUMERICAL RESULTS AND DISCUSSIONS}
\end{center}
\begin{sloppypar}
Since the densities stemming from the Dirac sea are evaluated
within the derivative expansion technique and expressed by means
of the mean fields as well as their derivative terms, the wave
functions of anti-nucleons are not involved when solving the
meson-field equations. The numerical procedure of the RHA is
similar to the one currently used in the RMF \cite{Rei86}, except
that one more equation for the anti-nucleon is implemented. It
goes as follows: in the n-th iteration step we have arrived at a
set of wave functions of nucleons and mean fields. First, we
calculate the densities contributed from the valence nucleons by
adding up the nucleon wave functions, and the densities originated
from the Dirac sea by evaluating the mean fields as well as their
derivative terms. Second, we determine the  meson fields by
solving the Laplace equations of mesons. Third, we use the new
meson fields to solve the Schr\"{o}dinger-equivalent equation of
the nucleon. Fourth, we compute the new single-particle energies
of nucleons and determine the occupation numbers by adjusting a
Fermi surface such that the particle number is conserved . This
completes one iteration step. The iteration is continued until the
binding energy is stable up to 6 digits. Finally, we apply the
known mean fields to solve the Schr\"{o}dinger-equivalent equation
of the anti-nucleon. The equation itself is solved iteratively.
The obtained wave functions are used to calculate the
single-particle energies of anti-nucleons. The space of
anti-nucleons are truncated by the specified principal and angular
quantum numbers $n$ and $j$ with the guarantee  that the
calculated single-particle energies of anti-nucleons are converged
when the truncated space is extended. We find that the results are
insensitive to the exact values of $n$ and $j$ provided large
enough numbers are given. We have used $n=4$, $j=9$ for $^{16}{\rm
O}$; $n=5$, $j=11$ for $^{40}{\rm Ca}$; and $n=9$, $j=19$ for
 $^{208}{\rm Pb}$.

\begin{center}
\fbox{Table I}
\end{center}

The parameters of the model are determined in a least-square fit
to the properties of spherical nuclei. The experimental values for
the observables used in the fit are given in Table I. The second
column gives the measured nuclear binding energies while the last
two columns reflect the properties of nuclear shape. In model
calculations one can extract the diffraction radius and surface
thickness through analyzing the nuclear charge form factor, where
the intrinsic nucleon form factors are buried in \cite{Rei86}.
Here we just want to note that instead of the commonly used Sachs
form factors \cite{Sim80} in the current fitting we have applied
recent parameterization of nucleon electromagnetic form factors
based on the Gari-Kr\"{u}mpelmann model \cite{Gar92,Lom01,Lom02}.
Specifically, we have taken the parameter set of GKex(02S)
presented in Ref. \cite{Lom02}.

In the fitting processes one can simply forget the anti-nucleon
part since the vacuum contributions to the densities are
calculated by means of the mean fields and their derivative terms.
Once the parameters are specified, we get a set of decided mean
fields which are then applied to solve the eigenequation of the
anti-nucleon. The eigenfunctions (the wave functions) and
eigenvalues (the single-particle energies) of anti-nucleons are
the final output of the model calculations. We intend, so to say,
to predict the bound states of anti-nucleons through adjusting the
model parameters to the bulk properties of finite nuclei.

Compared to the previous version of the RHA model now we have two
more parameters $f_{\omega}$ and $f_{\rho}$ for the tensor
couplings of vector mesons. The obtained parameters as well as the
corresponding saturation properties are presented in Table II and
denoted as the RHAT set. For the sake of comparison we give other
two sets of parameters: the NL1 set \cite{Rei86} of the RMF model
under the no-sea approximation and the RHA1 set \cite{Mao99} where
the vacuum contributions have been taken into account but without
tensor-coupling terms. One can see that after introducing the
tensor couplings a large effective nucleon mass remains in the RHA
model. The value of $m^{*}/M_{N} \approx 0.8$ is close to the one
appeared in the Skyrme-force parameterizations for nonrelativistic
approaches \cite{Fri86}. Although some rearrangements exist,
generally speaking, the changes of parameters between the RHAT set
and the RHA1 set are not very significant except that the vector
coupling strength ${\rm g}_{\omega}$ is enhanced. This causes a
somewhat larger compressibility.

\begin{center}
\fbox{Table II}  \hspace{6cm}  \fbox{Table III}
\end{center}

In Table III the values of $\chi^{2}$ are listed for the three
cases and the deviations from the binging energy, diffraction
radius and surface thickness are detailed. The spin-orbit
splitting in $^{16}{\rm O}$ and the shell fluctuation in
$^{208}{\rm Pb}$ are provided together with the empirical values.
Obviously, the NL1 set performs a very good fit and describes the
spin-orbit interaction satisfactorily. After including the vacuum
effects the RHA model still reaches a reasonable well fit to the
bulk properties of spherical nuclei. The effect of tensor-coupling
terms induces the spin-orbit force in the RHAT model to be one
times larger than that in the RHA1 model, and thus improves the
total $\chi^{2}$ value. In addition, one can observe an
interesting redistribution of deviations between the binding
energy and surface thickness. In the RHAT set the $\chi_{E}^{2}$
is suppressed substantially. It indicates an improved fit to the
nuclear binding energies as clearly exhibited in Table IV. In the
mean time, the contributions from the Dirac sea are enhanced
evidently.

\begin{center}
\fbox{Table IV}  \hspace{6cm}  \fbox{Fig. 1}
\end{center}

The shell fluctuation can typically be expressed via the charge
density in $^{208}{\rm Pb}$. It has been known that both the
relativistic and nonrelativistic mean-field theory overestimate
$\delta \rho$ by a factor of 3. From Table III one can find that
the NL1 set and the RHAT set share the same disease. However, the
RHA1 set reproduces the empirical value quite nicely. Further
investigation is needed in order to clarify whether it comes out
fortuitously. In Fig.~1 we depict the charge densities of three
spherical nuclei reckoned with the RHAT set of parameters. The
solid lines represent the experimental data. Except for the
amplitude of the shell fluctuations the overall results are in
agreement with the data.

\begin{center}
\fbox{Fig. 2}  \hspace{6cm}  \fbox{Fig. 3}
\end{center}

The contributions of the vacuum to the scalar density and baryon
density are shown in Fig.~2. The computations are performed with
the RHAT set of parameters for $^{40}{\rm Ca}$. Noticeable
influence from the Dirac sea can be found for the scalar density
while the effect on the baryon density is negligible. We have
solved the technical problem of fluctuations on the
$\rho_{S}^{sea}$ met in the previous RHA1 model through making
spline extrapolation for the first several points of densities
originated from the vacuum. Smooth curves for various densities in
different nuclei considered in this work have been obtained.
Fig.~3 shows the resultant scalar and vector potentials in
$^{16}{\rm O}$ for three models. Due to the vacuum effects the
potentials calculated with the RHA model are about half of that
computed with the RMF model. After introducing the tensor-coupling
terms, the RHAT set receives deeper potentials compared to the
RHA1 set, reflecting the effect of parameter rearrangements. The
enhancements are around 20 MeV for S and V in the center of the
nucleus, which are nonnegligible on the scale of the nucleon
central potential. Especially, the enhancements would be summed up
for the anti-nucleon potential rather than cancel each other for
the nucleon.

\begin{center}
\fbox{Table V}  \hspace{4cm}  \fbox{Table VI}  \hspace{4cm}
\fbox{Fig.~4}
\end{center}

In Table V and Table VI we present the single-particle energies of
protons (neutrons) and anti-protons (anti-neutrons) in three
spherical nuclei of $^{16}{\rm O}$, $^{40}{\rm Ca}$ and
$^{208}{\rm Pb}$. The binding energies per nucleon and the {\em
rms} charge radii are given too. The experimental data are taken
from Ref. \cite{Mat65}. Both the relativistic mean-field theory
(NL1) under the no-sea approximation and the relativistic Hartree
approach (RHA1, RHAT) taking into account the vacuum contributions
can reproduce the observed binding energies and {\em rms} charge
radii quite well. With respect to the large error bars in
measurements of the $1s$ proton (neutron) levels, the results of
all three sets of parameters coincide with the data. Because of
the large effective nucleon mass, the spin-orbit splitting on the
$1p$ levels is rather small in the RHA1 model. The situation has
been ameliorated conspicuously in the current RHAT model
incorporating the tensor couplings for the $\omega$- and
$\rho$-meson. At the same time, a large $m^{*}$ stays unchanged.
The experimental data for the anti-proton (anti-neutron) spectra
in the vacuum are presently unavailable. The RMF model and the RHA
model provide strikingly different predictions with a deviation of
a factor of 2, clearly demonstrating the importance of the Dirac
sea effects. On the other hand, the anti-particle energies
computed with the RHAT set of parameters are 20 -- 30 MeV larger
than that reckoned with the RHA1 set, as can be anticipated from
Fig.~3. The corresponding proton and anti-proton potentials in
$^{208}{\rm Pb}$ are displayed in Fig.~4. The same features
observed in the energy spectra are revealed once again.

\end{sloppypar}
\begin{center}
{\bf VI. SUMMARY AND OUTLOOK}
\end{center}
\begin{sloppypar}
We have incorporated tensor couplings for the $\omega$- and
$\rho$-meson in a relativistic Hartree approach for finite nuclei.
After refitting the parameters of the effective Lagrangian to the
bulk properties of spherical nuclei, the spin-orbit force has been
enlarged by a factor of 2 compared to the previous version of the
RHA model without tensor-coupling terms, while a large effective
nucleon mass remains. This improves the total $\chi^{2}$ value and
brings the computed proton (neutron) spectra more closer to the
data. The predicted anti-proton (anti-neutron) spectra in the
vacuum are deepened about 20 -- 30 MeV. One may argue that the
vacuum may not be properly treated by the nucleon degrees of
freedom, instead, a quark vacuum may be essential. However, if one
speaks about {\em observing} a vacuum, what one actually means is
to measure the response of the vacuum to the laboratory probes. In
the environment of a finite nucleus, nucleons are well established
physical degrees of freedom, which should be the relevant degrees
of freedom for describing the corresponding vacuum too. In the
case that a QCD environment is involved, quark degrees of freedom
may be necessary.

In view that in the vacuum nucleons and anti-nucleons are always
in the form of pairs, one of the promising ways to measure the
anti-nucleon spectra in the vacuum of a nucleus is to knock out
corresponding nucleons from the bound states emerging from the
lower continuum. The incident particles can be photons, electrons
or protons. The experimental searches for the vacuum structure
have been discussed in detail in Refs. \cite{Aue86,Jin88}. The
dynamical processes can be simulated by using the relativistic
Boltzmann-Uehling-Uhlenbeck approach \cite{Tei94,Ko87,Dan84,Mao94}
which is a microscopic transport model for single-particle
distribution functions. Since the nucleons excited from the bound
states of the Dirac sea have to overcome deep potentials in order
to become real particles, the final particle spectra as well as
the angular distributions should be different to that of nucleons
originated from the Fermi sea. Exclusive analyses of  observables
from the hadron-nucleus reactions at the energy range of several
GeV/c will exhibit the structure of quantum vacuum. Work on this
aspect is in progress. If the energy of the incident particle is
further increased, light nuclei could be directly excited from the
vacuum when the correlation effect is taken into account.
Relativistic quantum molecular dynamics model
\cite{Aic91,Ono96,Boh89,Bas98} is a suitable starting point to
study the relevant problems.

It is straightforward to extend the present model to include the
hyperon degrees of freedom. Then one can apply it to investigate
the properties of hypernuclei with the effects of quantum vacuum
taken into account. As a first step we consider the
single-$\Lambda$ and double-$\Lambda$ hypernuclei. Through
systematically studying the $\Lambda$ and nucleon spectra in the
Fermi sea and the anti-$\Lambda$ and anti-nucleon spectra in the
Dirac sea, one can extract important information for the hyperon
interaction, which is an active topic of modern nuclear physics
\cite{Sch94,Fil02,Hiy00,Ram97}. Here the anti-$\Lambda$ spectra in
the vacuum act as further constraints to the effective
interactions in addition to the usual considered hypernuclei
observables. The similar procedure can be performed for the $\Xi$-
and $\Sigma$-hypernuclei. The detailed understanding of the
hyperon-hyperon and hyperon-nucleon interactions in dense medium
is fundamental for the study of strange particle production and
strange particle flow in relativistic heavy-ion collisions
\cite{Ko97,Van99,Wan99} as well as the composition and structure
of neutron stars in astrophysics \cite{Gle85}.

\end{sloppypar}
 \begin{center}
{\bf ACKNOWLEDGMENTS}
 \end{center}
 \begin{sloppypar}
The author thanks P.-G.~Reinhard for stimulating discussions. This
work was supported by the National Natural Science Foundation of
China under the grant 10275072 and the Research Fund for Returned
Overseas Chinese Scholars.

 \end{sloppypar}

\newpage
\def\baselinestretch{1.0}
\begin{table}
\caption{The experimental values for the observables included in
the fit, the binding energy $E_{B}$, diffraction radius $R$ and
surface thickness $\sigma$. In the last line we also give the
adopted errors $\Delta {\rm O}_{n}$ for the fit.} \vspace{0.5cm}
\begin{center}
\begin{tabular}{lcccc}
\hline
\hline
                    &   $E_{B}$ (MeV)   &  $R$ (fm)   &  $\sigma$ (fm)  \\
 \hline
$^{16}{\rm O}$    &  $-$127.6  & 2.777       &   0.839   \\
$^{40}{\rm Ca}$   &  $-$342.1  & 3.845       &   0.978   \\
$^{48}{\rm Ca}$   &  $-$416.0  & 3.964       &   0.881   \\
$^{58}{\rm Ni}$   &  $-$506.5  & 4.356       &   0.911   \\
$^{90}{\rm Zr}$   &  $-$783.9  & 5.040       &   0.957   \\
$^{116}{\rm Sn}$  &  $-$988.7  & 5.537       &   0.947   \\
$^{124}{\rm Sn}$  &  $-$1050.0 & 5.640       &   0.908   \\
$^{208}{\rm Pb}$  &  $-$1636.4 & 6.806       &   0.900   \\    \\
$\Delta {\rm O}_{n} / {\rm O}_{n}$  &  0.2\%  &  0.5\%   &   1.5\% \\
\hline \hline
\end{tabular}
\end{center}
\end{table}

\begin{table}
\caption{Parameters of the RMF and the RHA models as well as the
corresponding saturation properties. $M_{N}$ and $m_{\rho}$ are
fixed during the fit.} \vspace{0.5cm}
\begin{center}
\begin{tabular}{lccc}   \\
 \hline \hline
 & NL1 & RHA1 & RHAT \\
 \hline
$M_{N}$ (MeV)        &  938.000 &  938.000  & 938.000 \\
$m_{\sigma}$ (MeV)   &  492.250 &  458.000  & 450.000 \\
$m_{\omega}$ (MeV)   &  795.359 &  816.508  & 814.592 \\
$m_{\rho}$ (MeV)     &  763.000 &  763.000  & 763.000 \\
${\rm g}_{\sigma}$   &  10.1377 &  7.1031   & 7.0899 \\
${\rm g}_{\omega}$   &  13.2846 &  8.8496   & 9.2215 \\
${\rm g}_{\rho}$     &  9.9514  &  10.2070  & 11.0023 \\
$b$ (fm$^{-1}$)      &  24.3448 &  24.0870  & 18.9782  \\
$c$                  &$-$217.5876 & $-$15.9936 & $-$ 27.6894 \\
$f_{\omega}/M_{N}$ (fm) & 0.0    &    0.0      & 2.0618   \\
$f_{\rho}/M_{N}$ (fm)   & 0.0    &    0.0      & 45.3318  \\
\\
$\rho_{0}$ (fm$^{-3}$)  &  0.1518   &   0.1524  &  0.1493  \\
$E/A$ (MeV)             &  $-$16.43 &  $-$16.98 & $-$ 16.76  \\
$m^{*}/M_{N}$           &  0.572    &   0.788   & 0.779  \\
$K$ (MeV)               &  212      &   294     & 311  \\
$a_{4}$ (MeV)           &  43.6     &   40.4    & 44.0  \\
\hline \hline
\end{tabular}
\end{center}
\end{table}

\begin{table}
\caption{The standard $\chi^2$ values of the parameterizations
given in Table II. The different sources of deviations from the
binding energy, diffraction radius and surface thickness are
separated. Other observables like the spin-orbit splitting of the
$1p$ level in $^{16}{\rm O}$ for both protons
($\delta\epsilon_{{\small P}}$) and neutrons
($\delta\epsilon_{{\small N}}$), and the shell fluctuation in
$^{208}{\rm Pb}$ ($\delta\rho$) are presented too.} \vspace{0.5cm}
\begin{center}
\begin{tabular}{lccccccc}   \\
 \hline \hline
 & $\chi_{E}^{2}$ & $\chi_{R}^{2}$ & $\chi_{\sigma}^{2}$ &
 $\chi_{tot}^{2}$ & $\delta\epsilon_{{\small P}}$ (MeV)&
 $\delta\epsilon_{{\small N}}$ (MeV)
 &   $\delta\rho$ (fm$^{-3}$)  \\
 \hline
 NL1 &  21.96  &  11.78  &  32.28  &  66.02  &  5.99  & 6.06 & $-$0.0070 \\
 RHA1 & 516.48 & 39.14   &  256.69 &  812.31 & 1.99  & 2.00 &$-$0.0030\\
 RHAT & 88.53 &  24.50  &   444.86 &  557.88 & 3.96 & 4.43 &$-$0.0067 \\
 Exp. &       &         &          &         & 5.98 & 6.07 &$-$0.0023\\
\hline \hline
\end{tabular}
\end{center}
\end{table}

\begin{table}
\caption{Experimental and theoretical binding energies per nucleon
within the RHA model. The vacuum corrections are indicated
explicitly.} \vspace{0.5cm}
\begin{center}
\begin{tabular}{lccccc}
\hline \hline
& & \multicolumn{2}{c}{RHA1} & \multicolumn{2}{c}{RHAT}    \\
& Exp. & Theory & Dirac Sea & Theory & Dirac Sea \\
 \hline
$^{16}{\rm O}$    &  $-$7.98  & $-$8.00  & 1.37  & $-$7.94 & 1.68 \\
$^{40}{\rm Ca}$   &  $-$8.55  & $-$8.73  & 1.43  & $-$8.62 & 1.74 \\
$^{48}{\rm Ca}$   &  $-$8.67  & $-$8.51  & 1.39  & $-$8.61 & 1.74 \\
$^{58}{\rm Ni}$   &  $-$8.73  & $-$8.44  & 1.44  & $-$8.62 & 1.85 \\
$^{90}{\rm Zr}$   &  $-$8.71  & $-$8.74  & 1.42  & $-$8.78 & 1.76 \\
$^{116}{\rm Sn}$  &  $-$8.52  & $-$8.61  & 1.39  & $-$8.52 & 1.68 \\
$^{124}{\rm Sn}$  &  $-$8.47  & $-$8.50  & 1.34  & $-$8.49 & 1.68 \\
$^{208}{\rm Pb}$  &  $-$7.87  & $-$7.93  & 1.30  & $-$7.88 & 1.62 \\
\hline \hline
\end{tabular}
\end{center}
\end{table}

\begin{table}
\caption{The single-particle energies of both protons and
anti-protons as well as the binding energies per nucleon and the
{\em rms} charge radii in $^{16}{\rm O}$, $^{40}{\rm Ca}$ and
 $^{208}{\rm Pb}$. } \vspace{0.5cm}
 \begin{center}
{\small
\begin{tabular}{ccccc}
\hline \hline
& NL1 & RHA1 & RHAT & Exp. \\
 \hline
$\;\;\;\;\;\;\;\;^{16}{\rm O}$ &       &        &      &       \\
$E/A$ (MeV)      &   8.00   & 8.00  & 7.94 &  7.98 \\
$r_{ch}$ (fm)    &   2.73   & 2.66  & 2.64 &  2.74 \\
 PROTONS         &          &       &      &       \\
$1s_{1/2}$ (MeV) &  36.18   & 30.68 & 31.63&  40$\pm$8 \\
$1p_{3/2}$ (MeV) &  17.31   & 15.23 & 16.18&  18.4   \\
$1p_{1/2}$ (MeV) &  11.32   & 13.24 & 12.22&  12.1   \\
 ANTI-PRO.       &          &       &      &         \\
$1\bar{s}_{1/2}$ (MeV) &  674.11   & 299.42 & 328.55 &     \\
$1\bar{p}_{3/2}$ (MeV) &  604.70   & 258.40 & 283.44 &     \\
$1\bar{p}_{1/2}$ (MeV) &  605.77   & 258.93 & 285.87 &     \\
\hline
$\;\;\;\;\;\;\;^{40}{\rm Ca}$ &         &      &      &       \\
$E/A$ (MeV)      &   8.58    & 8.73   &  8.62   &  8.55   \\
$r_{ch}$ (fm)    &   3.48    & 3.42   &  3.41   &  3.45   \\
PROTONS          &           &        &         &         \\
$1s_{1/2}$ (MeV) &  46.86    & 36.58  &  37.01  &50$\pm$11\\
$1p_{3/2}$ (MeV) &  30.15    & 25.32  &  25.95  &         \\
$1p_{1/2}$ (MeV) &  25.11    & 24.03  &  23.63  &34$\pm$6 \\
ANTI-PRO.        &           &        &         &         \\
$1\bar{s}_{1/2}$ (MeV)&796.09& 339.83 &  367.90 &         \\
$1\bar{p}_{3/2}$ (MeV)&706.36& 309.24 &  332.10 &         \\
$1\bar{p}_{1/2}$ (MeV)&707.86& 309.52 &  333.37 &         \\
\hline
$\;\;\;\;\;\;\;^{208}{\rm Pb}$ &      &         &      &       \\
$E/A$ (MeV)      &   7.89    & 7.93   &  7.88   &  7.87   \\
$r_{ch}$ (fm)    &   5.52    & 5.49   &  5.46   &  5.50   \\
 PROTONS         &           &        &         &         \\
$1s_{1/2}$ (MeV) &   50.41   & 40.80  &  41.74  &         \\
$1p_{3/2}$ (MeV) &   44.45   & 36.45  &  37.38  &         \\
$1p_{1/2}$ (MeV) &   43.75   & 36.21  &  37.18  &         \\
 ANTI-PRO.       &           &        &         &         \\
$1\bar{s}_{1/2}$ (MeV)&717.01&354.18  &  377.37 &         \\
$1\bar{p}_{3/2}$ (MeV)&705.20&344.48  &  366.95 &         \\
$1\bar{p}_{1/2}$ (MeV)&705.28&344.52  &  367.24 &         \\
\hline \hline
\end{tabular}
 } \end{center}
\end{table}

\begin{table}
\caption{The single-particle energies of both neutrons and
anti-neutrons.} \vspace{0.5cm}
\begin{center}
{\small
\begin{tabular}{ccccc}
\hline \hline
 & NL1 & RHA1 & RHAT & Exp. \\
 \hline
$\;\;\;\;\;\;\;\;^{16}{\rm O}$ &       &       &      &       \\
NEUTRONS          &          &         &       &            \\
$1s_{1/2}$ (MeV)  &  40.21   & 34.71   & 35.78 &  45.7   \\
$1p_{3/2}$ (MeV)  &  21.07   & 19.04   & 20.18 &  21.8   \\
$1p_{1/2}$ (MeV)  &  15.01   & 17.05   & 15.75 &  15.7   \\
ANTI-NEU.         &          &         &       &         \\
$1\bar{s}_{1/2}$ (MeV)&667.93& 293.23  & 322.47&         \\
$1\bar{p}_{3/2}$ (MeV)&598.74& 252.48  & 277.94&         \\
$1\bar{p}_{1/2}$ (MeV)&599.74& 252.97  & 279.22&         \\
\hline
$\;\;\;\;\;\;\;^{40}{\rm Ca}$  &       &       &      &       \\
NEUTRONS          &          &         &       &              \\
$1s_{1/2}$ (MeV)  &  54.85   & 44.48   & 44.98 &         \\
$1p_{3/2}$ (MeV)  &  37.79   & 32.98   & 33.83 &         \\
$1p_{1/2}$ (MeV)  &  32.73   & 31.71   & 30.99 &         \\
ANTI-NEU.         &          &         &       &             \\
$1\bar{s}_{1/2}$ (MeV)&783.87& 327.96  & 355.70&         \\
$1\bar{p}_{3/2}$ (MeV)&694.80& 298.04  & 321.07&         \\
$1\bar{p}_{1/2}$ (MeV)&696.18& 298.26  & 322.15&         \\
\hline
$\;\;\;\;\;\;\;^{208}{\rm Pb}$ &       &       &      &       \\
NEUTRONS        &             &        &       &         \\
$1s_{1/2}$ (MeV)  &  58.97   & 47.40   & 46.70 &         \\
$1p_{3/2}$ (MeV)  &  52.44   & 42.66   & 42.31 &         \\
$1p_{1/2}$ (MeV)  &  51.82   & 42.45   & 41.64 &         \\
ANTI-NEU.         &          &         &       &         \\
$1\bar{s}_{1/2}$ (MeV)&678.23& 313.18  & 334.39&         \\
$1\bar{p}_{3/2}$ (MeV)&667.70& 304.61  & 325.41&         \\
$1\bar{p}_{1/2}$ (MeV)&667.73& 304.61  & 325.28&         \\
\hline \hline
\end{tabular}
} \end{center}
\end{table}

 \newpage
 \begin{figure}[htbp]
 \vspace{-3cm}
 \mbox{\hskip -1.5cm \psfig{file=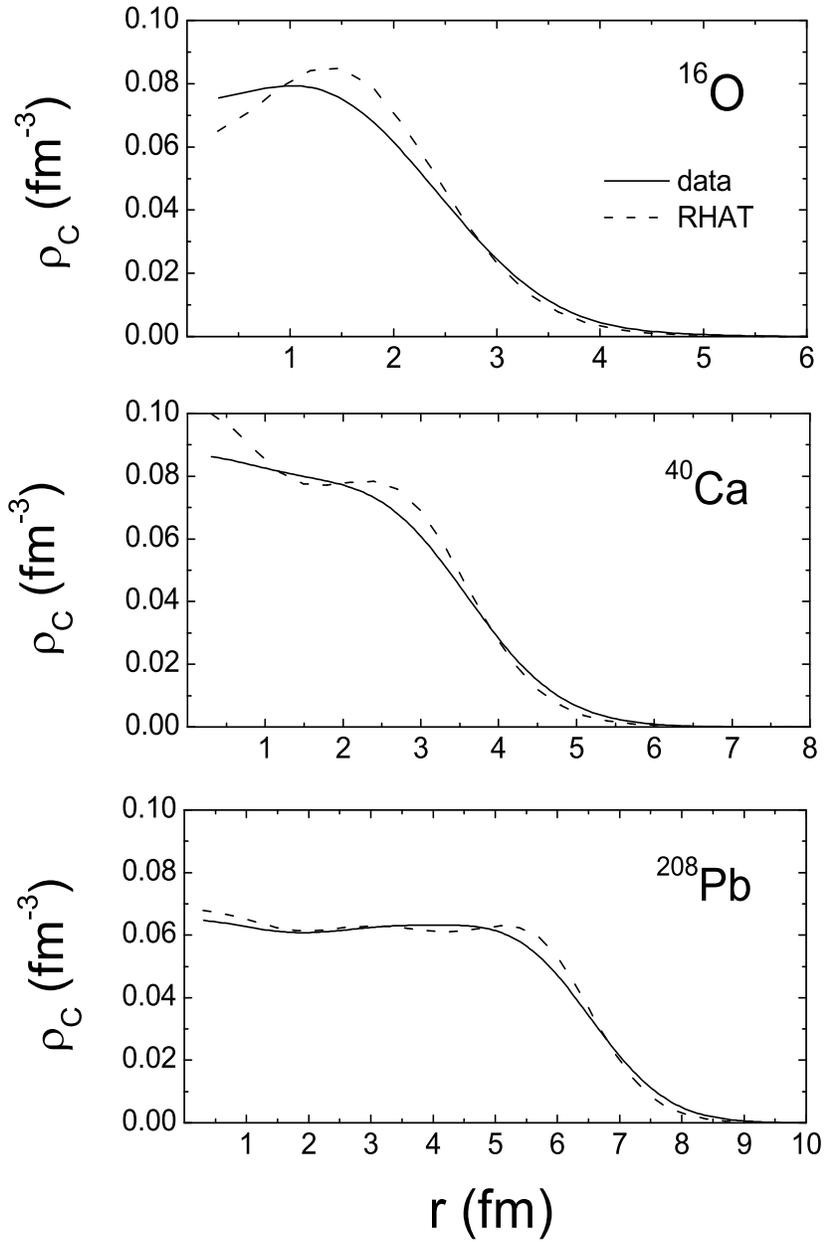,width=18cm,height=24cm,angle=0}}
 \vspace{-3.5cm}
 \caption{The charge densities in $^{16}{\rm O}$, $^{40}{\rm Ca}$
 and $^{208}{\rm Pb}$ computed with the RHAT set of parameters.
 The empirical data are depicted as solid lines.}
 \end{figure}

\newpage
 \begin{figure}[htbp]
 \vspace{-3.0cm}
 \mbox{\hskip -1.5cm \psfig{file=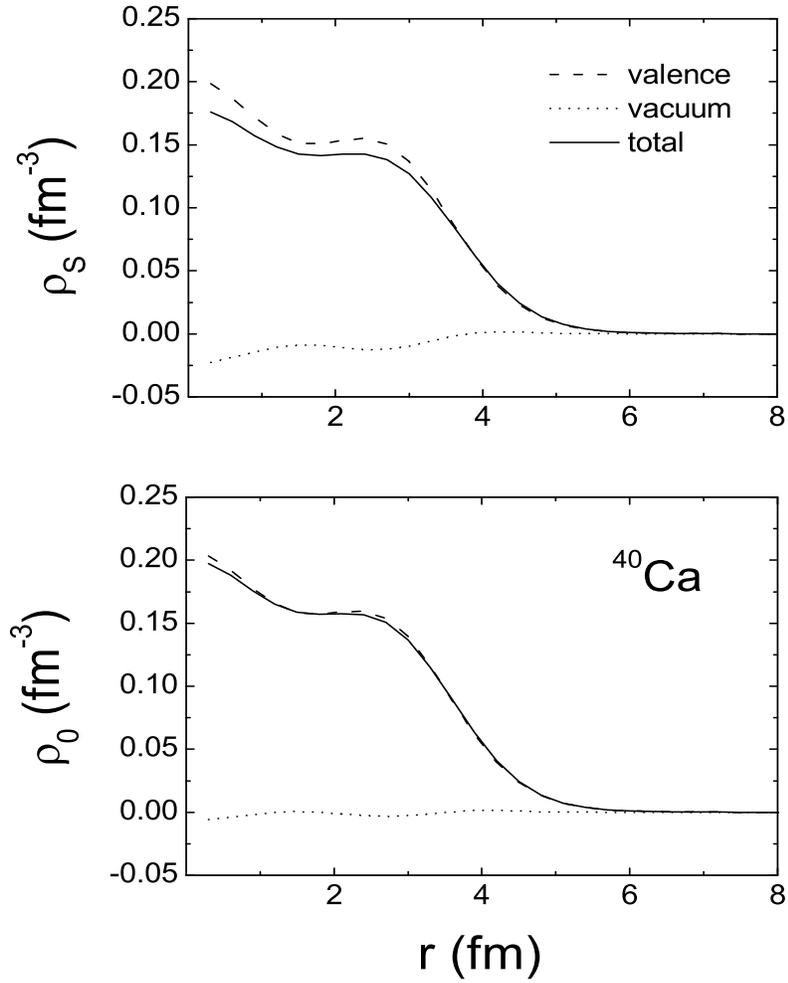,width=18cm,height=20cm,angle=0}}
 \vspace{-3.5cm}
 \caption{The scalar density and baryon density in $^{40}{\rm Ca}$. Dashed
 lines denote the contributions of valence nucleons, dotted lines represent
 the Dirac-sea effects and solid lines give the total results.}
 \end{figure}

 \newpage
 \begin{figure}[htbp]
 \vspace{-3.0cm}
 \mbox{\hskip -1.5cm \psfig{file=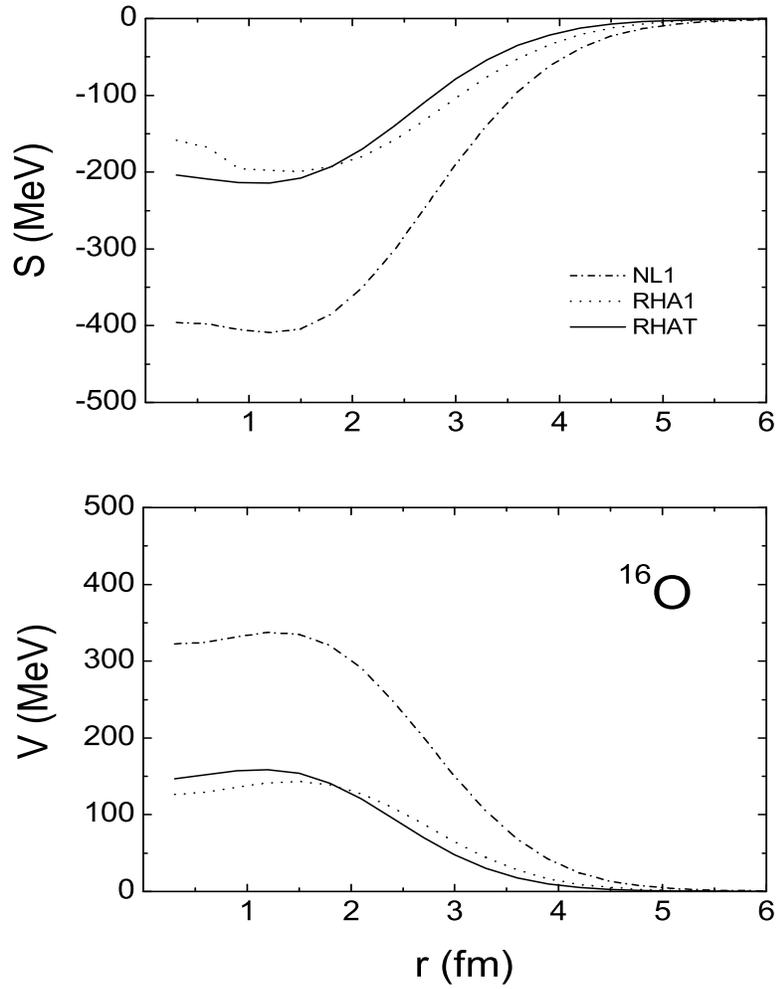,width=18cm,height=20cm,angle=0}}
 \vspace{-3.5cm}
 \caption{The scalar potential from the $\sigma$-meson exchange and the vector
 potential from the $\omega$-meson exchange in $^{16}{\rm O}$. Different curves
 are related to different sets of parameters as indicated in the figure.}
 \end{figure}

\newpage
 \begin{figure}[htbp]
 \vspace{-3.0cm}
 \mbox{\hskip 0.0cm \psfig{file=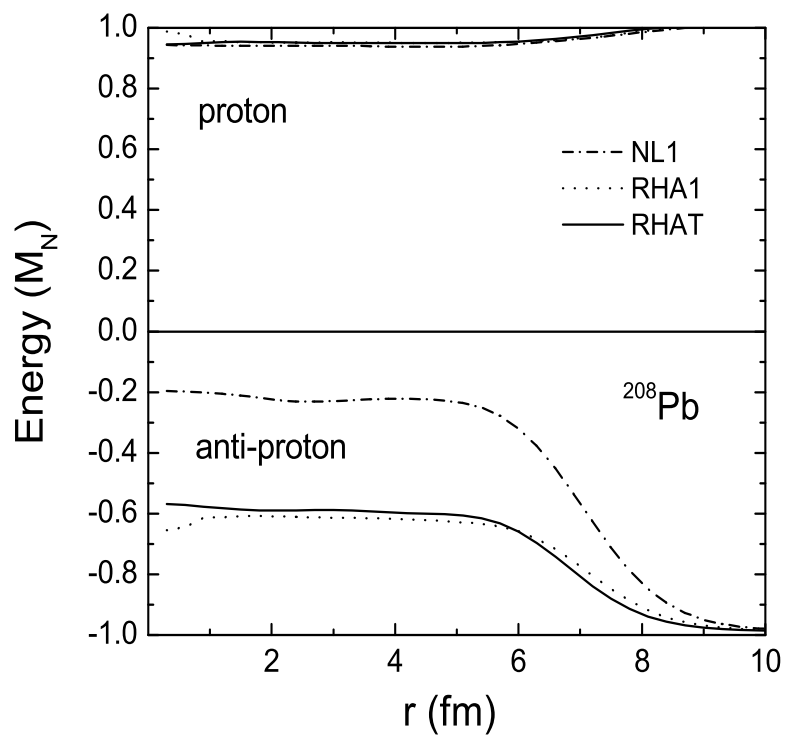,width=16cm,height=16cm,angle=0}}
 \vspace{-3.5cm}
 \caption{The potentials of the proton and the anti-proton in $^{208}{\rm Pb}$
 computed with different sets of parameters as indicated in the figure.}
 \end{figure}

\end{document}